\documentclass{ws-procs10x7}
\usepackage{balance}
\newlength{\ziffer}

\newcommand{\TeV}{\,\mbox{Te\kern-0.2exV}}
\newcommand{\GeV}{\,\mbox{Ge\kern-0.2exV}}
\newcommand{\mGeV}{\,\mathrm{Ge\kern-0.2exV}}
\newcommand{\MeV}{\,\mbox{Me\kern-0.2exV}}
\newcommand{\keV}{\,\mbox{ke\kern-0.2exV}}
\newcommand{\eV}{\,\mbox{e\kern-0.2exV}}

\newcommand{\ipb}{\,\mbox{pb}^{-1}}
\newcommand{\ifb}{\,\mbox{fb}^{-1}}

\newcommand{\bea}{\pagebreak[3]\begin{samepage}\begin{eqnarray}}
\newcommand{\eea}{\end{eqnarray}\end{samepage}\pagebreak[3]}
\newcommand{\beq}{\begin{equation}}
\newcommand{\eeq}{\end{equation}}

\newcommand{\abb}{Fig.~\ref}
\newcommand{\fig}{\abb}

 \newlength{\howlong}

\newcolumntype{d}[1]{D{.}{.}{#1}}

\makeindex
\begin{document}

\title{Top pair production III:\\Testing the Standard Model in Top Quark Decays}

\author{D. WICKE}

\address{
On behalf of the CDF and D\O\ collaborations\\~\\
Bergische Universit\"a{}t, Gau\ss{}str. 20, 42097 Wuppertal, Germany\\
E-mail: daniel.wicke@physik.uni-wuppertal.de}

\twocolumn[\maketitle\abstract{
With its discovery in 1995 by the CDF and D\O\ collaborations the top quark
completed the set of quarks expected by the Standard Model. It is predicted to
have the same quantum numbers and couplings as the other up-type
quarks. Albeit, only very few of these properties have been verified so far.
This article summarises the existing measurements of top quark properties in
the pair production mode.
}
]

\section{Introduction}\label{sect:intro}
The  set of quarks expected in the Standard Model (SM) of particle physics was
fully observed in 1995 with the discovery of the top quark by the CDF and D\O\ 
collaborations\cite{Abe:1995hr,Abachi:1995iq}.
All quantum numbers of the top quark except the mass are expected to
be identical to those of the other two up-type quarks. Within the Standard
Model these quantities fully define the production and decay properties of the top
quark.

Only few of these predicted properties have been verified to date.
These measurements were performed with the CDF and D\O\ experiments at the
Tevatron, which is still  the only accelerator available to study top quark physics.

This article  summarises the current status of measurements of top pair
decays which test the Standard Model predictions.
Measurements of single top production 
are covered separately\cite{wolfgang}.
The selection of top pairs and the measurement of the top quark mass 
are covered in separate articles in these proceedings\cite{dugan,florencia}.

Further reports on top quark results\cite{ChrisHill,AafkeKraan,StefanAnderson}
were given in the heavy quark session.

\section{$W$-Helicity}\label{sect:whel}
Within the Standard Model top quarks decay into a $W$-boson and a $b$-quark.
To check the expected $V\!-\!A$ structure of this weak decay the
$W$-helicity was investigated. 
Only left-handed particles are expected to  couple to the $W$-boson and thus
 the $W$
can be either left handed ($-$) or longitudinal ($0$). For the known $b$ and $t$ masses the
fractions should be $f_-\!=\!0.3$ and $f_0\!=\!0.7$, respectively. The right
handed (+) contribution is expected to be negligible.

Depending on the $W$-helicity ($-,0,+$) the charged lepton in the $W$ decay prefers
to align with the $b$ direction, stay orthogonal or escape in the direction
opposite to the $b$.

Several observables are sensitive to the helicity: the  transverse
momentum of the lepton, $p_T^\mathrm{lept}$, the lepton-$b$-quark invariant mass, $M_{lb}$
and the angle between the lepton and the $b$-quark direction. For best
sensitivity at Tevatron energies this angle is measured in the $W$ rest
frame.

Currently the best results are obtained using $M_{lb}^2$  and
the angle between lepton and $b$-quark, $\cos \theta^*$. 
Unless otherwise stated the results assume $f_0\!=\!0.7$ which remains unchanged
for a $V\!+\!A$ admixture.
%They shall be described in the following.

\subsection{Results using  $M_{lb}^2$ }
CDF has released an analysis\cite{CDF2006iy} based on the lepton-$b$-quark invariant mass
using both the lepton+jets channel with ${\cal L}=695\ipb$ and the dilepton channel
with ${\cal L}=750\ipb$.

\begin{figure}[b]
\centerline{\includegraphics[width=0.4\textwidth]{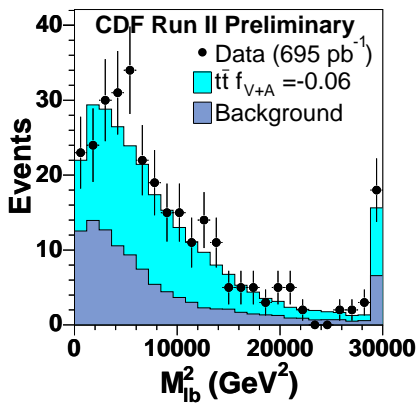}}
\caption{\label{fig:cdfmlb}%
$M_{lb}^2$ distribution for lepton+jets data compared to simulation.
}
\end{figure}

In the lepton+jets channel one measurement per event can be
performed. $b$-tagging is required in the event selection. For
events with only one identified $b$-jet the invariant mass of the identified
jet with the identified lepton is used to create a $M_{lb}^2$ distribution,
see \fig{fig:cdfmlb}. 
For events with two identified $b$-jets $M_{lb}^2$ is computed for both jets
and used to create a 2 dimensional (2-d) distribution.

%The obtaind data distributions are then compared to $V\!-\!A$- and $V\!+\!A$-templates
%obtained from simulation including signal and background processes.

Dilepton events
%$b$-tagging doesn't help to reduce the ambiguity in
%assigning jets and thus isn't required. Each 
provide two simultaneous measurements.
Both possible pairings of leptons to jets are used to
create a 2-d $M_{lb}^2$ distribution with two entries per event.

%Again 
The obtained data distributions are compared to  $V\!-\!A$ and 
$V\!+\!A$ templates obtained from simulation including signal and background
processes.

A binned $\log$-likelihood fit procedure is used to extract the fraction of
$V\!+\!A$ in data in the three samples. The final 
left-handed $W$ fraction is
\bea
f_+&=&-0.02\pm0.07\nonumber\\
f_+&<&0.09\quad 95\%\mbox{CL}%\mathrm{.}
\eea
in agreement with the SM expectation.
\subsection{Results using  $\cos\theta^*$ }
Both CDF and D\O\ used  $\cos\theta^*$  to measure the $W$-helicity. 

\subsubsection{D\O}
\label{d0costheta}
D\O\ investigated about $370\ipb$ in both the dilepton and the lepton+jets
channel\cite{Abazov:2006hb}. 

In the lepton+jets channel $\cos \theta^*$ is reconstructed from the measured
objects after improving the resolution with a constraint fit using 
the $W$ and top masses. Only the jet assignment of the best fit
is considered.
%
%% \begin{figure}
%% \includegraphics[width=0.4\textwidth,clip]{figs/T27F02_color.eps}\\
%% \caption{\label{fig:d0l+jetscostheta}%
%% d0l+jetscostheta
%% }
%% \end{figure}

\begin{figure}[b]
\vspace*{-5mm}
%\centerline{\includegraphics[width=0.4\textwidth]{figs/T27F04c.eps}}
\centerline{%
\includegraphics[width=0.40\textwidth]{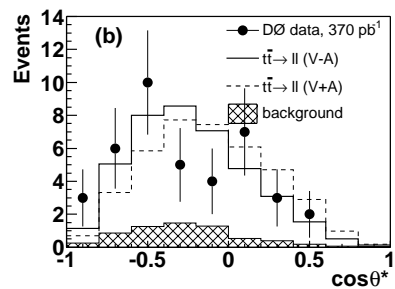}%
}
\caption{\label{fig:d0costhetadilepton}%
$\cos\theta^*$  measured in dilepton events.
}
\end{figure}

In the dilepton channel the same constraints are used to reconstruct the neutrino
momenta with a fourfold ambiguity. 
In order to account for detector resolution the recontruction is repeated 100
times with object momenta smeared according to the experimental resolution.
Both possible assignments of the leading 2 jets to the leptons are considered.
The result from the repetitions and assignments are averaged to obtain two
$\cos\theta^*$ measurements per event (\fig{fig:d0costhetadilepton}).
%
%% \begin{figure}
%% \includegraphics[width=0.4\textwidth,clip]{figs/T27F03_color.eps}\\
%% \caption{\label{fig:d0dilepcostheta}%
%% d0dilecostheta
%% }
%% \end{figure}

The resulting $\cos\theta^*$ distributions are compared to $V\!+\!A$ and $V\!-\!A$
templates from simulation that include background contributions. To obtain the
final result a maximum likelihood fit is used.
D\O\ obtains:
\bea
f_+&=&0.056\pm0.080\pm0.057\nonumber\\
f_+&<&0.23\quad 95\%\mbox{CL}\mathrm{.}
\eea
consitent with the SM prediction.\\~

\subsubsection{CDF}
CDF has performed two different analyses in the lepton+jets channel using
${\cal L}=1\ifb$.

 The template based method\cite{CdfNote8368} is very similar to the D\O\ analysis
described in Sect.~\ref{d0costheta}. \fig{fig:cdftemplate} shows the data
distribution and the fitted template curves which result in:
\bea
f_+&=&-0.05\pm0.06\pm0.03\nonumber\\
f_+&<&0.11\quad 95\%\mbox{CL}\mathrm{.}
\eea
\begin{figure}\vspace*{-5mm}
\centerline{\includegraphics[width=0.4\textwidth]{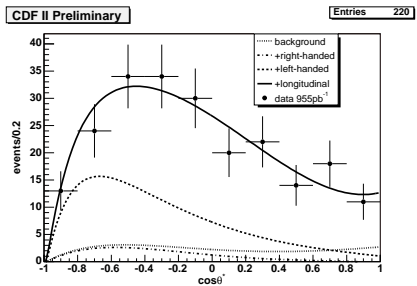}}
\caption{\label{fig:cdftemplate}%
Distribution of $\cos\theta^*$ measured by CDF compared to template predictions.
}
\end{figure}

 The second analysis\cite{CdfNote8380} uses migration matrices to transfer the theory prediction
to the level of measured quantities.
It also uses all possible jet assignment weighted according
to their signal probability based on kinematic properties. It obtains
\bea
f_+&=&-0.03\pm0.03\pm0.04\nonumber\\
f_+&<&0.10\quad 95\%\mbox{CL}\mathrm{.}
\eea

\begin{figure}\vspace*{-5mm}
\centerline{\includegraphics[width=0.4\textwidth,clip,trim=0mm 7mm 0mm 14mm]{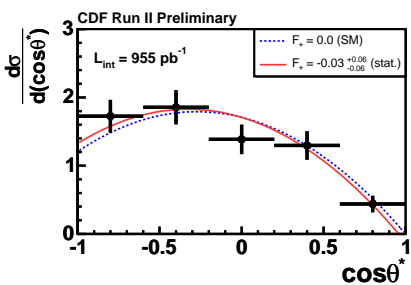}}%
\caption{\label{fig:cdfolded}%
Unfolded $\cos\theta^*$ distribution by CDF compared to Standard Model and best fit theory.
}
\end{figure}

Both CDF measurements are also reinterpreted as measurements of $f_0$ assuming the
SM contribution of $f_+$. Both results are compatible with the SM of
$f_0\simeq 70\%$
within their 20\% uncertainty.

\section{Top Charge}\label{sect:charge}
The top quarks electrical properties are fixed by its charge. 
However, in reconstructing top quarks the charges of the objects 
usually aren't checked. Thus an exotic charge
value of $\left|q_t\right|=4e/3$ isn't excluded by standard analyses.

D\O\ performed a reconstruction of the top charge in the lepton+jets channel\cite{Abazov:2006vd}
using ${\cal L}=370\ipb$. Two or more identified $b$-jets are required. The
assigment of $b$-jets to the leptonic or hadronic event side is based on the
quality of a fit to the $t\bar t$ hypothesis, which uses the $W$ and top
masses as constraints.

The charge is then reconstructed for both top quarks in the event as
\bea
Q_\mathrm{lep}&=&\left|q_l+q_{b_\mathrm{l}}\right|\nonumber\\
Q_\mathrm{had}&=&\left|-q_l+q_{b_\mathrm{h}}\right|\mbox{,}
\eea
where the charge of the $b$-jets in the leptonic and hadronic side,
$q_{b_\mathrm{l}}$ and $q_{b_\mathrm{h}}$, are obtain using the charged tracks
assigned to the $b$-jets.
%\begin{figure}
%\includegraphics[width=0.4\textwidth]{figs/T06DF01.eps}%
%\caption{\label{fig:d0jetcharge}%
%d0 jetcharge
%}
%\end{figure}

The reconstructed top charge distribution are compared to templates from
simulation of the two cases and judged with unbinned likelihoods.
From the likelihood ratio D\O\ excludes the $4e/3$ case at
92\% CL and limits its fraction in the current data sample to less than 80\% at
90\% CL.
\begin{figure}[b]\vspace*{-4.5mm}
\centerline{\includegraphics[width=0.35\textwidth]{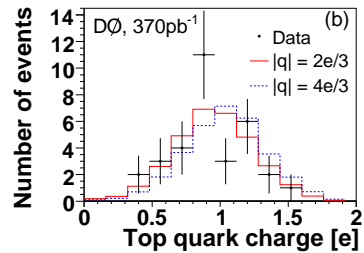}\vspace*{-2mm}}
\caption{\label{fig:d0charge}%
Top charge in D\O\ data and simulation.
}
\end{figure}
%\begin{figure}
%\includegraphics[width=0.4\textwidth]{figs/T06DF03.eps}%
%\caption{\label{fig:d0chargelikelihood}%
%d0 charge likelihood
%}
%\end{figure}
%
%
%Paper: hep-ex/0608044
%Date: Wed, 16 Aug 2006 21:35:28 GMT   (24kb)
%
%Title: First determination of the electric charge of the top quark
%Authors: D0 Collaboration, V. Abazov, et al
%Comments: 7 pages, 3 Postscript figures, submitted to Phys. Rev. Lett
%Report-no: FERMILAB-PUB-06/278-E

\section{Top Branching Ratio}\label{sect:topbr}
In top quark decays any down-type quark can be
produced in association with the $W$-boson. 
The relative strength of these modes is governed by the 
CKM matrix. With an expected $\left|V_{tb}\right|\simeq 0.999$ the decay into
$b$-quarks is by far dominating in the Standard Model.
Various extension of the Standard Model allow for deviations from
these expectations.

In top pair production only the branching ratio
% of $t\rightarrow W+b$ over $t\rightarrow W+$any quark
\bea
 R = \frac{B(t\rightarrow Wb)}{B(t\rightarrow Wq)}
=\frac{\left|V_{tb}\right|^2}{\left|V_{td}\right|^2\!\!+\!\left|V_{ts}\right|^2\!\!+\!\left|V_{tb}\right|^2}\quad
\eea
is experimentally accessible.
CDF and D\O\ have performed  analyses 
%on $162\ipb$ and $230\ipb$, respectively%
that determine $R$ from the 
ratio of $t\bar t$ events with two, one or no 
$b$-tags\cite{Acosta:2005hr,Abazov:2006bh}:
\bea
\begin{array}{rclcrcl}
\multicolumn{3}{l}{\mbox{D\O~}(230\ipb)}&&
\multicolumn{3}{l}{\mbox{CDF~}(162\ipb)}
\\[0.5em]
R &=& 1.03^{+0.19}_{-0.17}& ~\quad\, &R &=& 1.12^{+0.27}_{-0.23}\\
R &>& 0.64                && R &>& 0.62 \quad 95\%\mbox{~CL.}\\
\end{array}
\eea
%% \begin{figure}
%% \centerline{\includegraphics[height=4cm]{figs/T06AF01a_color.eps}}
%% \caption{\label{fig:d0vtb}D\O's 2-d fit to the number of selected events and
%%   the ratio of branching ratios $R$.
%% }
%% \end{figure}

These results are regularly converted to limits on the CKM matrix element
\beq
\left|V_{tb}\right|>0.80  ~\qquad \left|V_{tb}\right|>0.78 \quad 95\%\mbox{~CL.}\\
\eeq
It is important to note that in these analyses a sensitivity to deviations 
from the expected $\left|V_{tb}\right|\simeq 1$ would only be
reached if the fraction of top quarks that decays to
$b$-quarks was comparable to the light quark fractions.
%\begin{figure}
%{\setlength{\unitlength}{0.5mm}
%\begin{picture}(1,1)
%\put(105,1){\tiny$R_{\mathrm{rec}}$}
%\put(5,70){\tiny$R$}
%\end{picture}%
%\includegraphics[height=4cm]{figs/FC_sys2.eps}
%}
%\caption{\label{fig:cdfvtb}asdfasdf
%}
%\end{figure}

\section{Searches for New Particles}\label{sect:searches}
Further results are obtained by checking the %explicit 
presence of new
particles in $t\bar t$ decays. 

CDF has searched for $H^+$ in top decays by checking the
cross-sections of various Standard Model modes including one $\tau$ channel
and looking for modes that are expected in $H^+$ decays. Limits on the
branching fraction $B(t\rightarrow bH^+)$ are set for various $H^+$ 
decay modes and masses\cite{Abulencia:2005jd}.

From the reconstructed top mass distribution CDF infers
the possible mass of a $t^\prime$ particle to be
$M_{t^\prime}>258\GeV$ at 95\% CL\cite{CdfNote8495}.

CDF and D\O\ have investigated the distribution of the invariant mass of
$t\bar t$ pairs and checked for resonances. Cross-section limits for narrow
resonances are obtained and presented as mass limits in a %reference 
$Z^\prime$ model\cite{CdfNote8087,d0note4880}.

\section{Summary}\label{sect:summary}
Several properties of the top quark are investigated in $t\bar t$ events by
the CDF and D\O\ collaborations. 

Measurements of the $W$-Helicity can probe 10\% $V\!+\!A$ contributions, the
excotic charge value of the top quark can be excluded at 92\%~CL and
measurements of the decay flavour have reached 20\% precision.
More analyses including
searches for unknown particles in $t\bar t$ events have been performed.

No significant deviation from the Standard Model has been observed (yet).
\bibliographystyle{utphysnt}
\begin{flushleft}
\bibliography{InTheseProc,CDF,Dzero}
\end{flushleft}

\end{document}
%%% Local Variables: 
%%% mode: latex
%%% TeX-master: t
%%% End: 